\documentstyle[12pt]{article}
\date{}
\begin{document}

\title{The Charm and Bottom Hyperons in a Chiral Quark Model}

\author{L. Ya. Glozman$^{1,2,3}$ and D. O. Riska$^3$}
\maketitle

\centerline{\it $^1$Institute for Theoretical Physics,
University of Graz, 8010 Graz, Austria}

\centerline{\it $^2$Alma-Ata Power Engineering Institute, 
480013 Alma-Ata, Kazakhstan}

\centerline{\it $^3$Department of Physics, University of Helsinki,
00014 Finland}

\setcounter{page} {0}
\vspace{1cm}

\centerline{\bf Abstract}
\vspace{0.5cm}

The spectrum of the $C=1$ hyperons is well
described by the constituent quark model, if the fine structure
interaction between the light and strange
quarks is mediated by the $SU(3)_F$ octet of light
pseudoscalar mesons, which are the Goldstone bosons of the hidden
approximate chiral symmetry of QCD. With the addition of a 
phenomenological flavor exchange interaction of the same form
between the light and the charm quarks to describe the 
$\Sigma_c-\Sigma_c^*$ and $\Xi_c^s-\Xi_c^*$ 
splittings,
the splittings between the 
$C=1$ states fall within 10-30 MeV of the empirical values. Predictions
are presented for the lowest negative parity excited states and
the magnetic moments as well. Corresponding predictions for the
$B=-1$ hyperon states are also given.\\
\\

Preprint HU--TFT--95--53, hep-ph 9509269

\newpage

\centerline{\bf 1. Introduction}
\vspace{0.5cm}

A very satisfactory description of the
whole observed part of the spectrum of the nucleon, $\Delta$-resonance
and the strange hyperons can be achieved with the constituent quark
model if in addition to an
effective harmonic confining interaction the constituent quarks 
are assumed
to interact by exchange of the
$SU(3)_F$ octet of light pseudoscalar mesons, which are the
Goldstone bosons of the hidden realization of the approximate chiral
symmetry of QCD [1,2].
We here show that the extension of this model
to the ground state spectrum of the $C=1$ charm hyperons predicts a
spectrum in good agreement with the empirically known states
if it is augmented by a
weaker phenomenological heavy flavor exchange interaction 
that acts between
the light and the charm and the strange and the charm quarks 
respectively.
This flavor exchange interaction may be viewed as arising from
exchange of $D$ (and $D^*$) mesons (or systems with the same
quantum numbers) between the
$u,d$ and the $c$-quarks and of $D_s$ (and $D_s^*$) mesons
between the $s$ and $c$ quarks. Such flavor exchange interactions 
requires complete
antisymmetrization of the 3 quark states, even for the quarks of
widely different mass. This implies a close formal correspondence
between the symmetry (and notation) for the $C=+1$ and $S=-1$ 
hyperon states [3,4].\\

The mean energy of the 2 lowest excited $\Lambda_c^+$ states is
predicted to be in good agreement with the empirical value, under the
assumption that these states are the charm analogs of the
strange flavor singlet $\Lambda(1405)-\Lambda(1520)$ negative parity 
resonances. 
Together with the satisfactory prediction of the energies of the
$C=+1$ hyperons in the ground state band this indicates that
the chiral constituent quark model of refs.[1,2] has
the proper heavy quark limit. The predicted spectrum of the ground
state bottom $B=-1$ hyperons is very similar to that of the $C=+1$
charm hyperons. Predictions are given for the lowest ("P-shell")
negative parity states as well as for the magnetic moments,
with inclusion of the small exchange current
corrections. \\

The spin-spin component of the $SU(3)_F$
pseudoscalar octet exchange interaction
has the form 

$$H_{\chi}=-\sum_{i<j}\{\sum_{a=1}^{3}V_\pi (r_{ij})\lambda_i^a
\lambda_j^a+\sum_{a=4}^{7} V_K(r_{ij})\lambda_i^a \lambda_j^a$$
$$+V_\eta (r_{ij})\lambda_i^8\lambda_j^8\}\vec \sigma_i \cdot \vec
\sigma_j,\eqno(1.1)$$
where the radial functions $V_{\pi}$, $V_K$ and $V_\eta$ represent the
$\pi$, $K$ and $\eta$-exchange interactions respectively, and which have the
usual Yukawa form at long range. As the behavior of these functions at
short range is not known, their matrix elements in the lowest shells
of the harmonic oscillator basis were extracted from the lowest
splittings of the nucleon spectrum in ref. [2]. \\

With those matrix elements the spectra of the light and strange
baryons were predicted to be in remarkably good agreement with the empirical
spectra. Moreover the chiral boson exchange interaction (1) leads to
the correct ordering of the positive and negative parity states in the
spectra in the spectrum and in contrast to the commonly employed gluon
exchange model [5]. The motivation for the interaction (1) is 
the unique role that the light pseudoscalar octet mesons have
as Goldstone bosons of the spontaneously broken approximate chiral
symmetry of QCD. Because of
their large masses the corresponding charm and charm-strange
pseudoscalar mesons $D$ and $D_s$ cannot on the other hand 
be given any such interpretation. In fact by the near degeneracy 
between them and the corresponding
vector mesons $D^*$ and $D_s^*$ a meson exchange interaction of the
form (1) would be expected to be due to both
pseudoscalar and vector meson exchange. A phenomenological interaction
that represents these types of interaction mechanisms in pairs involving
one light and one heavy quark would have the form 

$$H_h=-\sum_{i<j}\{\sum_{a=9}^{12}V_{D}(r_{ij})\lambda_i^a
\lambda_j^a$$
$$+\sum_{a=13}^{14}V_{D_s}(r_{ij})\lambda_i^a\lambda_j^a\}
\vec \sigma_i \cdot \vec
\sigma_j.\eqno(1.2)$$
Here the $\lambda_i^a$ matrices are the $SU(4)$ extension of the
$SU(3)_F$ Gell-Mann matrices in (1). As the quark content of the
$\eta_c$ and the $J/\psi$ are purely $c\bar c$ these mesons
mediate no
interaction between the light and charm quarks, and are therefore
omitted from
(1.2). 
It will be shown below that an interaction of this form is
required to explain the nonvanishing
$\Sigma_c^*-\Sigma_c$ and $\Xi_c^*-\Xi_c^s$ splittings.
The interactions $V_{D}$ and $V_{D_s}$ will be treated as
purely phenomenological here, and in fact we shall only need their
$S$-state matrix elements  in the oscillator basis.\\

As the interaction (1) does not act in quark pair states that 
include charm or bottom quarks, the
consequence is that the fine structure of the $C=+1$ and $B=-1$
hyperons should be mainly determined by the interaction (1) in the
quark pair with only light and/or strange quarks. As moreover the
ground state band is insensitive to the details of the confining
interaction, this implies that the energies of the ground state charm
and bottom flavor hyperons may, to a first approximation, be 
predicted without any additional
parameters.\\

This paper is devided into 7 sections. In section 2 we
review the harmonic oscillator basis states for the hyperons with 
one heavy quark, treating the quark mass difference as a
perturbation. In
section 3 we construct the ground state spectra of the charm and
bottom hyperons using the chiral field interaction (1.1) and discuss
the role of the flavor exchange interactions between light and heavy
quarks. In section 4 we discuss the lowest excited negative parity
states of the charm hyperons. Section 5 contains the results for the
ground and $L=1$ state bands of the $B=-1$ hyperons. 
In section 6 we give predictions for the magnetic
moments of the ground state $C=+1$ and $B=-1$ hyperons. Section 7
contains a summarizing discussion.\\

\vspace{1cm}

\centerline{\bf 2. The Basis States for the Heavy Flavor Hyperons}
\vspace{0.5cm}

There is no a priori reason to exclude two-body flavor exchange
interactions between constituent quarks. Such are automatically
implied by any direct quark couplings to the light pseudoscalar mesons, 
once the meson degrees of freedom are integrated out of the
corresponding Fock space. The importance of such couplings
appears increasingly compelling
[2,6,7,8]. An immediate consequence of flavor exchange interactions
is that such
imply the necessity of complete antisymmetrization of the
3 quark states that form the baryons even when the 3 quarks have
different flavors and constituent masses. This is most
readily illustrated by an example. Consider the $\Lambda^0$, the
quark content of which is $uds$. By definition the $u, d$ quark
pair
state has to be completely antisymmetric. Kaon exchange between
say the $u$ and the $s$ quarks will exchange their flavor coordinates
thus leading to an antisymmetric $d,s$ pair state. Iteration of the
argument implies that 
the three quarks have to be in antisymmetric flavor states, and as
a corollary that
the $uds$ state is necessarily totally
antisymmetric. The argument generalizes immediately to the
$\Lambda_c^+$, which has the quark content $udc$. In the latter case
the flavor exchange interaction could be
mediated e.g. by the $D$ or $D^*$
mesons. \\

The requirement of the total antisymmetry of the 3-quark
wavefunctions, independently of the quark flavors, is met by
the $SU(3)$ flavor-spin 
basis for the charm quarks, which can be constructed
directly from the corresponding 3-quark states for the
strange hyperons [3,9] by replacing the $s$ quark with the
$c$ quark in the $\Lambda$ and $\Sigma$ flavor
states in the case of the $\Lambda_c^+$ and the 
$\Sigma_c$, by replacing either the $u$ and the $s$ or the
$d$ and the $s$ quark in the $\Lambda$ and $\Sigma$
flavor states by the $s$ and $c$ quarks respectively
in case of the $\Xi_c^a$ and $\Xi_c^s$ and finally by replacing
the light quark by the $c$ quark in the $\Xi$ wavefunctions in the case 
of the
$\Omega_c^0$.
The only new feature in the case
of the $C=+1$ and
$B=-1$ hyperons is the appearance of the states $\Xi_c^{a}$ and 
$\Xi_b^{a}$ in
which the light and strange quarks form an antisymmetrical
combination, as well as the appearance of spin $1/2$ $\Omega_c^0$ and
$\Omega_b^-$ states in addition to the spin $3/2$ $\Omega_c^{0*}$ and
$\Omega_b^{-*}$ states, which have the same flavor-spin symmetry as
the $\Omega^-$ [3,4].\\

We shall describe the effective confining interaction by a 
harmonic oscillator interaction with flavor independent string
tension. The harmonic oscillator Hamiltonian for the 3-quark
system is then

$$H_0=\sum_{i>1}^{3}{\vec p_i^2\over 2m_i}-{\vec
P_{cm}^2\over 2 M}+{1\over
6}\sum_{i<j}k(\vec r_i-\vec r_j)^2,\eqno(2.1)$$
where $M$ is the sum of the 3 quark masses $(\sum_{i=1}^{3}m_i)$, and
$\vec P_{cm}$ is their total momentum. 
In the case of the $C=1$ or $B=-1$ hyperons two of the quark
masses represent the constituent masses of the light or strange
quarks ($m$) and one that of a heavy flavor quark ($m_h$).\\

Because
of the mass difference between the light and heavy quarks,
the antisymmetrization of the 3 quark wavefunction will lead to a
mixing of the orbital and flavor states so that the hyperon states
will lack definite orbital and flavor symmetry. 
In view of the fact that the harmonic oscillator 
potential represents but a crude effective representation of the
confining interaction we shall here avoid the
diagonalization of the oscillator Hamiltonian in the antisymmetric
basis with mixed orbital and flavor symmetry and be content to
treat the
quark mass difference as a flavor dependent perturbation to the
equal mass model Hamiltonian:

$$H_0^{'}=\sum_{i>1}^{3}{\vec p_i^2\over 2m}-{\vec
P_{cm}^{2}\over 6m}+{1\over
6}\sum_{i<j}m\omega^2(\vec r_i-\vec r_j)^2,\eqno(2.2)$$
where $\omega=\sqrt{k/m}$. The perturbation that arises from the
quark mass differerence is then 

$$H_0^{''}=-\sum_{i=1}^3\frac{m_h-m}{2m}\lbrace
\frac{\vec p_i^2}{m_h}-\frac{\vec P_{cm}^2}{3(2m+m_h)}\rbrace
\delta_{ih}.\eqno(2.3)$$
Here the Kronecker $\delta_{ih}$ indicates that the perturbation
acts only when i equals the coordinate label of the heavy quark.
If the term that contains the center-of-mass momentum 
$\vec P_{cm}$ is dropped in (2.3) the only effect of the
perturbation (2.3) on the oscillator states in the ground state 
band will be a lowering of their energies by
$$<g.s.\vert H_0^{''}\vert g.s.> = -\frac{1}{2}\delta, \eqno (2.4)$$
where

$$\delta=(1-m/m_h)\omega.\eqno(2.5)$$
The flavor-spin symmetries of the hyperons in the ground state band 
are $[3]_{FS}[21]_F[21]_S$ and $[3]_{FS} [3]_F [3]_S$, where $[f]_i$
denotes a Young pattern with $f$ being the sequence of integers that
indicates the number of boxes in the successive rows of the Young
pattern. This should then be combined with the totally symmetric
orbital state ($[3]_X$) and the antisymmetric color state
($[111]_C$).\\

The flavor-spin symmetries of the zero order oscillator wavefunctions
of the lowest lying negative parity
$\Lambda_c^+$ ($\Lambda_b^0)$ states in the $P$-shell will be
$[21]_{FS}[111]_F[21]_S$, 
$[21]_{FS}[21]_F[21]_S$  and
$[21]_{FS}[21]_F[3]_S$ respectively. The corresponding zero order
wavefunctions of the
$\Sigma_c(\Sigma_b)$ states will have the flavor-spin symmetries
$[21]_{FS}[21]_F[21]_S$, $[21]_{FS}[3]_F[21]_S$ and
$[21]_{FS}[21]_F[3]_S$. The levels of all of these states, which are
degenerate at zero order (2.2), will be split by the chiral field
interaction (1.1), the heavy flavor exchange interaction (1.2), as
well as by the perturbation (2.3) that arises from the quark mass
difference. The flavor-spin symmetry of the predicted negative parity
resonances of the $\Xi_c^a$ and the $\Xi_c^s$ are the same as these of
the $\Lambda_c$ and the $\Sigma_c$ above. The states and their
symmetry assignments are listed in Tables 1-3.\\

\vspace{1cm} 

\centerline{\bf 3. Fine Structure Splitting of the $C=1$ Ground
State Hyperons.}
\vspace{0.5cm}

The spectrum of the ground state $C=1$ charm hyperons is
obtained by treating the flavor exchange interactions (1.1)
and (1.2) in first order perturbation theory.  
The correction to the
unperturbed level is then expressed in terms of S-state
matrix elements of the potential functions $V_\pi, V_K$ and
$V_\eta$ in the pseudoscalar octet exchange interaction (1.1),
as well as of the charm and charm-strangeness exchange
potentials in $H_H$ (1.2). We shall denote these matrix
elements $P_{nl}^f=<nlm\vert V\vert nlm>$ , where the superscript
$f$ indicates the type of flavor exchange. 
Here $\vert nlm>$ are 3-dimensional harmonic oscillator
wavefunctions. In the case of
$\eta$ exchange the constituent masses of the quarks in the 
interacting
pair is indicated explicitly (here $m_u=m_d$).
The fine structure
corrections to the different hyperon states are expressed in
terms of such integrals in Table 1. In these fine structure
corrections we have also included the difference $\Delta_s$ 
between
the constituent masses of the light $u, d$ and $s$ quarks, as well as
the energy shift $-\frac{1}{2}\delta$ (2.4) that is
caused by the quark mass difference.\\

	The pion and $K$ exchange matrix elements $P_{00}^\pi$
and $P_{00}^K$ were extracted from the empirical
$\Delta_{33}-N$ and $\Sigma(1385)-\Sigma$ mass differences
to be 29.05 MeV and 20.1 MeV in ref. [2]. The $u-s$ quark mass
difference was determined from the $\Lambda^0-N$ mass difference
to be 127 MeV. Finally the matrix element $P_{00}^{us}$
of the $\eta$ exchange potential in $u,s$ and $d,s$ quark
pair states was assumed to equal the matrix element of the
$K$-exchange interaction: $P_{00}^{us}$=$P_{00}^K$. The
two remaining matrix elements of the $\eta$ exchange
interaction for pair states of light quarks $P_{00}^{uu}$
and of $s$-quarks $P_{00}^{ss}$ were determined from the 
matrix element $P_{00}^{us}$ using the quark mass scaling
relations

$$P_{nl}^{us}=(\frac{m_u}{m_s}) P_{nl}^{uu},
\quad P_{nl}^{ss}=(\frac{m_u}{m_s}) P_{nl}^{us}.\eqno(3.1)$$
The constituent masses of the $u$ and $d$ quarks were
taken to be equal and to be 340 MeV, and hence $m_s=467$
MeV.\\

We shall treat the matrix elements of the charm ($D$)
and strangeness-charm ($D_s$) exchange interaction
potentials in (1.2) as phenomenological parameters, denoted
$P_{00}^{D}=<000|V_D(r)|000>$ and $P_{00}^{D_s }=
<000|V_{D_s}(r)|000>$ respectively.
In view of the
near degeneracy of the charm and the charm-strange
$D$ and $D_s$ pseudoscalar
mesons and of the corresponding $D^*$ and $D_s^*$ vector
mesons we shall assume the matrix element equality

$$P_{00}^{D}= P_{00}^{D_s}.\eqno (3.2)$$
It will be shown below that the empirical 
$\Sigma_c^*$--$\Sigma_c$ splitting indicates the magnitude 
of these matrix elements to be about 3 times smaller than that 
of the corresponding $K$ exchange matrix element $P_{00}^K$.\\
	
The numerical predictions for the $C=1$ ground state
charm hyperon mass values are given in Table 1. The masses of all
these states, with the exception of the $\Omega_C^{0*}$, have
now been determined experimentally [10,11,12]. In the Table
we give the predicted mass values with and without the
phenomenological $D$ and $D_s$ interaction matrix
elements, which are required in the present model for lifting
the degeneracy between the $\Sigma_c$ and $\Sigma_c^*$ 
and that between the $\Xi_c^s$ and $\Xi_c^*$.\\

	The magnitude of the matrix element $P_{00}^{D}$
may be extracted from the empirical 
splitting between the $\Sigma_c$
and the $\Lambda_c^+$:

$$m(\Sigma_c)-m(\Lambda_c^+)=8P_{00}^\pi-4P_{00}^D
-\frac{4}{3}P_{00}^{uu}.\eqno(3.3)$$
This yields the value $P_{00}^{D}=6.5$ MeV. It is worth
noting that the ratio between this value and the corresponding
value 20.1 MeV for the $K$ exchange interaction matrix element
$P_{00}^K$ is close to the quark mass ratio $m_s/m_c$ that
would be suggested by comparison of the expressions for the
$K$ and $D$ exchange pseudoscalar exchange interactions, if the
coupling strengths of these two interactions are equal. To see
this we note that the value for the constituent mass of the
$s$-quark was determined to be 467 MeV in [2]. The corresponding
value for the $c$-quark may be determined from the difference
betweent the $\Lambda_c^+$ and $\Lambda^0$:

$$m_c=m(\Lambda_c^+)-m(\Lambda^0)+m_s+6P_{00}^K-6P_{00}^{D}
-{1\over 2}\delta.\eqno(3.4)$$
This yields the value $m_c=1652$ MeV, when in the expression for
the mass difference correction $\delta$ (2.5) we use the value
$\hbar\omega=$ 157 MeV [2]. 
These values for $m_s$ and $m_c$ are very close to those
obtained in ref.[13].
The ratio $m_s/m_c=0.28$,
is then only slightly smaller than the matrix element ratio
$P_{00}^{D}/P_{00}^K=0.32$. The fact that the latter number is
slightly larger is natural, as the relative importance of the
vector meson exchange interaction should be larger in the case
$D$ and $D^*$ exchange than in the case of $K$ and $K^*$
exchange in view of the near degeneracy of the $D$ and $D^*$ mesons.
The near equality between the quark 
mass and matrix element ratios
suggests that 
the overall interaction strength
is approximately $SU(4)_F$ symmetric and that
this flavor symmetry is broken mainly through the
quark mass differences. \\

With the numerical value 6.5 MeV for $P_{00}^D$ the 
$\Sigma_c$ mass is fitted to be in agreement with the
experimental value, whereas if the $D$ exchange matrix
element is set to 0, the mass of the $\Sigma_c$ is overpredicted
by 26 MeV. Since this represents only $\simeq$ 15 \% of the mass 
splitting between the $\Sigma_c$ and the $\Lambda_c^+$
it is clear that the dominant part of the hyperfine splitting
of the hyperons with only one heavy quark is due to the 
hyperfine interaction between the two light quarks. This
value for $P_{00}^D$ is however not sufficient to
explain all of the splitting between the $\Sigma_c^*$ and
the $\Sigma_c$, which is solely due the fine structure interaction
between the light and the charm quarks. This splitting
is obtained as $6 P_{00}^D=$ 39 MeV, which is considerably
smaller than the empirical splitting of 75 MeV. As the present
experimental mass value for the $\Sigma_c^*$ remains very
uncertain there is little reason at this time to increase the
value for $P_{00}^D$ to reduce this difference, most
of which (60 MeV)
could be accounted for by taking $P_{00}^D$ to
be 10 MeV, at a price of a concomitant (quite insignificant)
underprediction of 14 MeV of the $\Sigma_c$.\\

The empirical 
splitting between the $\Xi_c^a$ and the $\Sigma_c$ is only
10 -- 15 MeV. The expression for this splitting is in the
present model

$$m(\Xi_c^a)-m(\Sigma_c)=P_{00}^\pi-6P_{00}^K
+\frac{1}{3}P_{00}^{uu}-2P_{00}^{us}+7P_{00}^{D}
-3P_{00}^{D_s}+\Delta_s.\eqno(3.5)$$

With the matrix element values above this splitting is also
predicted to be small: 30 MeV.
The agreement with the empirical
splitting may be improved by exploiting the at
least 10 MeV large uncertainty in the value 127 MeV for the
quark mass difference parameter $\Delta_s$.
Perfect agreement with the empirical
splitting can of course be achieved by relaxing the
assumed matrix element equality (3.2) and taking
$P_{00}^{D_s}$ to be 11 MeV rather than 6.5 MeV.
The smallness of the $\Xi_c^a-\Sigma_c$ splitting does in fact
provide a rather sensitive test of the model, in view
of the requirement of balancing the $u-s$ quark mass
difference against the fine structure matrix elements.\\

The empirical splitting between the $\Xi_c^s$ and the
$\Xi_c^a$ is 95 MeV, within a considerable uncertainty
range. The present prediction for this splitting is

$$m(\Xi_c^s)-m(\Xi_c^a)=4P_{00}^K
+\frac{8}{3}P_{00}^{us} -2P_{00}^{D}
-2P_{00}^{D_s}, \eqno(3.6)$$
the numerical value of which is 108 MeV. This is only
about 10\% larger than the empirical  splitting 90-95 MeV.\\

The splitting between the $\Xi_c^*$ and the $\Xi_c^s$
is predicted to be
$$m(\Xi_c^*)-m(\Xi_c^s)=3P_{00}^{D}+3P_{00}^{D_s}.
\eqno(3.7)$$
Numerically this is 39 MeV, which amounts to only about
one half of the empirical splitting 82 MeV. Because of the
substantial uncertainty in the empirical mass value for the
$\Xi_c^s$ the predicted value may nevertheless turn out
to be satisfactory. The splitting can of course in principle
be fully accounted for
in the same way as the $\Sigma^*-\Sigma_c$
splitting by increasing the values of the matrix
elements $P_{00}^{D}$ and $P_{00}^{D_s}$.\\

The splitting between the $\Omega_c^0$ and the $\Xi_c^a$
is predicted to be

$$m(\Omega_c^0)-m(\Xi_c^a)=6P_{00}^K+P_{00}^{us}
-\frac{4}{3}P_{00}^{ss}+3P_{00}^{D}+3P_{00}^{D_s}
+\Delta_s.
\eqno(3.8)$$
The numerical value for this splitting is 241 MeV, which
agrees very well with the empirical value 245 MeV. Finally
the $\Omega_c^{0*}-\Omega_c^0$ splitting is predicted to be
$6P_{00}^{D_s}$ = 39 MeV.\\

It is worth noting that the fine structure corrections
in Table 1 imply the equal spacing rule [14]
$$m(\Sigma_c^*)-m(\Sigma_c)=m(\Xi_c^*)-m(\Xi_c^s)
=m(\Omega_c^{0*})-m(\Omega_c^0) \eqno (3.9)$$
under the matrix element equality assumption (3.2).
With the near equality of the present empirical values for 
the first two
of these splittings (75 MeV, 82 MeV) this rule appears to
be well satisfied.
This equal spacing rule implies the weaker mass relation
for these charm hyperons proposed in ref. [15].\\

It is interesting to note how similar the present predictions of the
splittings of the ground state charm hyperons are to those obtained
previously in the topological soliton (Skyrme) model, which implies 
an underlying
large $N_c$ limit. The $\Sigma_c-\Lambda_c^+$ splitting was predicted to
be 154-170 MeV in ref. [16] and 179 MeV in ref. [17]. These values are
close to the present predictions in Table 1 and to the empirical one.
The quark model based prediction of the splitting between the
$\Sigma_c^*$ and the $\Sigma_c$ above of $\sim$39 MeV falls within the
range 38-62 MeV predicted in ref. [16], but is larger than the value
25 MeV obtained in ref. [17].\\

\vspace{1cm}

\centerline{\bf 4. The Negative Parity States with $L=1$}
\vspace{0.5cm}

The unperturbed harmonic oscillator energies for the lowest negative
parity excitations with $L=1$ have equal energy and lie $\hbar \omega$
above the unperturbed ground state.  The corrections to this
unperturbed level that arise from the flavor exchange interactions
(1.1) and (1.2) are readily calculated using the methods of ref. [2],
and are listed in Tables 2 and 3 for the $\Lambda_c^+$,
$\Sigma_c$ and $\Xi_c$ and $\Omega_c^0$ respectively. In these
expressions we also have included the correction that arises from mass
difference perturbation (2.3) in lowest order. This perturbation is
flavor dependent and takes the following values in the different
$P$-shell multiplets:

$$<\Lambda_c^+|H_0^{''}|\Lambda_c^+>_{[21]_{FS}[111]_F[21]_S}=-{2\over
3}\delta \eqno(4.1a)$$
$$<\Lambda_c^+|H_0^{''}|\Lambda_c^+>_{[21]_{FS}[21]_F[21]_S}=-{2\over
3}\delta \eqno(4.1b)$$
$$<\Lambda_c^+|H_0^{''}|\Lambda_c^+>_{[21]_{FS}[21]_F[3]_S}=-{7\over
12}\delta.\eqno(4.1c)$$
The corresponding corrections for the $P$-shell excitations of the
$\Sigma_c$ are

$$<\Sigma_c^+|H_0^{''}|\Sigma_c^+>_{[21]_{FS}[21]_F[21]_S}=-{2\over
3}\delta,\eqno(4.2a)$$
$$<\Sigma_c^+|H_0^{''}|\Sigma_c^+>_{[21]_{FS}[3]_F[21]_S}=-{2\over
3}\delta,\eqno(4.2b)$$
$$<\Sigma_c^+|H_0^{''}|\Sigma_c^+>_{[21]_{FS}[21]_F[3]_S}=-{3\over
4}\delta.\eqno(4.2c)$$
Finally the corresponding corrections for the $L=1$ negative parity
excitations of the $\Xi_c^a$ are the same as those of the
$\Lambda_c^+$ (4.1), and those of the $\Xi_c^b$ are the same as those
of the $\Sigma_c$ (4.2) (neglecting the difference between the $u,d$
and $s$ quarks in this correction).\\

In view of the short range of the heavy flavor exchange interaction it
is natural to expect the $P$-shell matrix elements of this
interaction to be small, as the corresponding oscillator
wavefunctions vanish at short range. With $P_{11}^{D}=0$
the predicted energy of the central of the lowest negative parity
$\Lambda_c^+$ multiplet falls at 2599 MeV, which is only 10 MeV
below the corresponding empirical value 2609 MeV. The latter value is
obtained under the assumption that the two recently discovered
$\Lambda_c^+\, (2593)$ and $\Lambda_c^+\, (2625)$ resonances form a
negative parity spin doublet, which corresponds to the low lying
$\Lambda (1405)-\Lambda (1520)$ 
strange flavor singlet spin doublet. The
small underprediction of 10 MeV can in principle be removed by chosing
$P_{11}^{D}$ = 5 MeV. We shall accordingly employ this value in the
numerical predictions below, although the smallness of
this value makes the
changes from the predictions that are obtained by setting it to 0 are
in fact too small to be significant at the expected level of accuracy 
of the model.\\

The predicted energies of the other $\Lambda_c^+$ and $\Sigma_c$
negative parity states in Table 2 are expected to be fairly realistic
in view of the remarkably satisfactory prediction obtained for the
centroid energy of the $\Lambda_c^+(2593)-\Lambda_c^+(2625)$ doublet.
The present prediction for splitting between the ground state and
this doublet is 25 - 50 MeV larger than the corresponding ones
obtained in refs.[18,19], in which the fine structure interaction
between the quarks was described in terms of one gluon exchange.\\

In Table 3 we list the predicted energies of the negative parity
excited states of the $\Xi_c$ and the $\Omega_c^0$ hyperons
with with $L=1$. We expect the reliability of these predictions
to be similar to those obtained for the corresponding states of
the $\Lambda_c$ and $\Sigma_c$ hyperons above.\\

Among the negative parity states listed in Tables 2 and 3 are notes
that some are predicted to be near degenerate. Thus the $\Lambda_c^+$
multiplets with zero order
wavefunctions with
mixed flavor symmetry $[21]_F$ 
are predicted to be split by only 12 MeV. A similar near
degeneracy is predicted for the $[21]_F$ $\Xi_c^a$ with $S=1/2$ and
$[21]_F$ $\Xi_c^s$ with $S=3/2$. The ${1\over 2}^-$ and ${3\over 2}^-$
of these doublets may therefore be too close to be experimentally
resolvable.\\

The present prediction of the position of the energy of the
$\Lambda_c(2593)^+-\Lambda_c(2625)^+$ negative parity doublet is the
first that is in agreement with the empirical value. The quark
model predictions of ref. [5] overpredicts the position by 70 MeV, and
that of ref. [18] overpredicts the spin-orbit splitting of the
multiplet by a large factor. The soliton model prediction of ref. [16]
also overpredicts this spin-orbit splitting.\\

\vspace{1cm}

\centerline{\bf 5. The Spectrum of the $B=-1$ Hyperons}
\vspace{0.5cm}

The spectrum of the $B=-1$ hyperons that is predicted with the same
model as that used above for the $C=+1$ hyperons will differ from the
former in only two aspects. The first is that the larger constituent
mass of the $b$-quark will increase the mass difference correction
(2.5) slightly and the second is the fact that the generalization of
the heavy flavor exchange interaction (1.2) to the case of the bottom
hyperons should be less important than for the charm hyperons 
in view of the very short range of
bottom flavor exchange mechanisms and the smallness of the overall
factor 1/$m_b$ that is expected to be associated with the
$B$-meson exchange interaction. The latter of these two features would
imply that the splitting between the $\Sigma_b$ and $\Sigma_b^*$, as
well as $\Xi_b^{(s)}$ and $\Xi_b^*$ and the $\Omega_b^-$ and
$\Omega_b^{-*}$ states should be very small (no more than $\sim$ 10-20
MeV). The first data on the masses of the $\Sigma_b$ and $\Sigma_b^*$
hyperons does however give their mass splitting as 56 MeV, although
within a large uncertainty limit [20]. This - if confirmed - indicates
that $B$-exchange mechanisms cannot be neglected.\\

The expression for the mass difference between the $\Sigma_b$ and
$\Lambda_b$ is, in analogy with (3.3), 

$$m(\Sigma_b)-m(\Lambda_b)=8P_{00}^\pi-4P_{00}^B-{4\over
3}P_{00}^{uu}.\eqno(5.1)$$
Here $P_{00}^{B}$ is defined as the matrix element $<000|V_B(r)|000>$,
where $V_B(r)$ is the effective $B$-meson exchange interaction in the
$SU(4)_F$ extension of the interaction (1.2). With the 
$\Sigma_b-\Lambda_b^0$ mass difference
value 173 MeV [20] and using the same values for the matrix elements
of the $\pi$ and $\eta$ exchange interactions we then obtain
$P_{00}^B=$ 6 MeV, which is almost as large as the corresponding value
for the $C$-exchange matrix element $P_{00}^{D}$.\\

The value for the constituent mass of the $b$-quark may be determined
from the mass splitting between the $\Lambda_b^0$ mass and the
$\Lambda^0$ in analogy with (3.4) as

$$m_b=m(\Lambda_b^0)-m(\Lambda^0)+m_s+6P_{00}^K-6P_{00}^B-{1\over
2}\delta.\eqno(5.2)$$
This gives $m_b$ = 5003 MeV. Here we take $m_h=m_b$ in the expression
(2.5) for the quark mass difference correction. \\

The predicted $B=-1$ hyperon states in the ground state band are
listed in Table 4. The predicted $\Sigma_b^*-\Sigma_b$ splitting is
$6P_{00}^B=36 MeV$ if $P_{00}^B$ is taken to be 6 MeV as suggested by
the empirical $\Sigma_b-\Lambda_b^0$ splitting. This splitting falls
within the uncertainly limits of the empirical splitting $56\pm 22$
[20]. The predicted value for the $\Xi_b^a-\Sigma_b$ splitting is

$$m(\Xi_b^a)-m(\Sigma_b)=P_{00}^\pi-6P_{00}^K+{1\over
3}P_{00}^{uu}-2P_{00}^{us}+7P_{00}^B-3P_{00}^{B_s}
+\Delta_s.\eqno(5.3)$$
If the matrix element of the $B_s$ exchange interaction 
$P_{00}^{B_s}=<000\vert V_{B_s}(r)\vert 000>$
is taken to be equal to that of the $B$-exchange 
interaction (6 MeV) this
splitting comes out as 28 MeV. The splitting between the $B=-1$
cascade hyperons $\Xi_b^s$ and $\Xi_b^a$ is 

$$m(\Xi_b^s)-m(\Xi_b^a)=4P_{00}^K+{8\over
3}P_{00}^{uu}-2P_{00}^B-2P_{00}^{B_s}.\eqno(5.4)$$
Using the same matrix element values as above, this splitting is
predicted to be 110 MeV.\\

The splitting between the $\Xi_b^*$ and the $\Xi_b^s$ is predicted (in
analogy with (3.7)) to be $3P_{00}^B+3P_{00}^{B_s}=$ 36 MeV. The
splitting between $\Omega_b^-$ and the $\Xi_b^a$ is

$$m(\Omega_b^-)-m(\Xi_b^a)=6P_{00}^K+P_{00}^{us}-{4\over
3}P_{00}^{ss}+3P_{00}^B+3P_{00}^{B_s}+\Delta_s.\eqno(5.5)$$
The predicted numerical value for this splitting is 238 MeV - i.e. it
should be almost equal to that between the $\Omega_c^0$ and the
$\Xi_c^a$. Finally the $\Omega_b^{*-}-\Omega_b^-$ splitting is
predicted to be $6P_{00}^{B_s}\simeq$ 36 MeV. The predicted mass
values in Table 4 are close to those obtained in ref. [13], once the
latter are shifted up by the 20 MeV needed to bring the predicted mass
of the $\Lambda_b$ into agreement with its empirical value.\\

The size of the empirical splitting between the $\Sigma_b^*$ and the
$\Sigma_b$ is only 2 times smaller than that between the $\Sigma_c^*$
and the $\Sigma_c$. This is larger than the ratio $\simeq 1/3$ that
would be suggested by both the present pseudoscalar exchange model,
and the gluon exchange model for the hyperfine interaction. The
empirical splitting is also much larger than the prediction of the
bound state version of the Skyrme model [16,17].\\

In Table 5 we list the predicted negative parity excitations with
$L=1$ of the $B=-1$ hyperons. In these predictions we have not
included any $B$-meson exchange interactions, as the contribution
from such are expected to be smaller than the uncertainty
range of the predictions obtained with the 
present model and as in the absence of empirical data on the
energies of the $B=-1$ hyperon resonances in the $N=1$ band
the p-state matrix elements of the $B-$ and $B_s$-exchange
interactions are unknown. Because of
the neglect of bottom exchange interactions the $S=1/2$ multiplets 
within the different quark flavor combinations are predicted to be 
degenerate.\\

The predicted central position of the lowest $\Lambda_b^0$ negative
parity multiplet in Table 5 is 300 MeV above the $\Lambda_b^0$. This is
$\sim$30 MeV below the corresponding predictions obtained in ref. [5],
where the fine structure interaction between the quarks was described
in terms of gluon exchange. The Skyrme model predictions of ref. [16]
for this central position are 30-50 MeV lower.\\

\vspace{1cm}

\centerline{\bf 6. Magnetic Moments of the Heavy Flavor Hyperons}
\vspace{0.5cm}

The expressions for the magnetic moments of the ground state charm
hyperons in the constituent quark model in the impulse
approximation have been derived in ref. [21]. These expressions, which
are linear combinations of ratios of the nucleon and relevant
constituent quark masses, are listed in Table 6. In the table the
corresponding numerical values are also listed as obtained with the
constituent mass values used above (i.e. $m_u=$ 340 meV, $m_s=$ 467
MeV, $m_c=$ 1652 MeV).\\

A flavor dependent interaction between the quarks of the form (1.1) or
(1.2) implies charge exchange between quarks of unequal charge, and
hence also of two-body or exchange current operators [2,22]. The
general form of the octet vector exchange current operator that is
associated with the pseudoscalar octet mediated interaction (1.1) will
have the form [2]: 

$$\vec \mu^{\pi K}=\mu_N\{\tilde{V}_\pi(r_{ij})(\vec \tau_i\times \vec
\tau_j)_3$$
$$+\tilde{V}_K(r_{ij})(\lambda_i^4\lambda_j^5-\lambda_j^4\lambda_i^5)\}(\vec
 \sigma_i\times \vec \sigma_j).\eqno(6.1)$$
Here $\tilde{V}_\pi(r)$ and $\tilde{V}_K(r)$ are dimensionless
functions, which describe the spatial structure of the $\pi$ and $K$
exchange magnetic moment
operators and $\mu_N$ is the nuclear magneton.\\

The charm exchange interaction (6.2) will give rise to a similar
exchange current operator that involves $SU(4)_F$ matrices with the form 

$$\vec
\mu^{DD_s}=\mu_N\{\tilde{V}_D(r_{ij})(\lambda_i^{11}\lambda_j^{12}
-\lambda_i^{12}\lambda_j^{11})$$
$$+\tilde{V}_{D_s}(r_{ij})(\lambda_i^{13}\lambda_j^{14}
-\lambda_i^{14}\lambda_j^{13})\}(\vec
 \sigma_i\times \vec \sigma_j)\eqno(6.2)$$
Here $\tilde{V}_D(r_{ij})$ and $\tilde{V}_{D_s}(r_{ij})$ are then the
corresponding dimensionless functions, which describe the spatial
structure of the $D$ (or $D$ and $D^*$) and $D_s$ (or $D_s$ and
$D_s^*$) exchange magnetic moments.\\

The exchange corrections to the magnetic moments of the hyperons in
the ground state band will only depend on the $S$-shell matrix
elements of the spatial functions $\tilde{V}_a(r)$ in (6.1) and (6.2)
($a=\pi,K,D,D_s$).\\

\noindent
The expressions for these exchange current corrections to the magnetic
moments of $J={1\over 2}$ the charm hyperons are then

$$\mu^{ex}(\Lambda_c^+)=-\mu^{ex}(\Sigma_c^+)=-{1\over
2}\mu^{ex}(\Sigma_c^0)$$
$$=2<\varphi_{000}(\vec
r_{12})|\tilde{V}_D(r_{12})|\varphi_{000}(\vec
r_{12})>\mu_N,\eqno(6.3a)$$
$$\mu^{ex}(\Xi_c^{a+})=-\mu(\Xi_c^{s+})=-\mu(\Omega_c^0)=$$
$$2<\varphi_{000}(\vec
r_{12})|\tilde{V}_{D_s}(r_{12})|\varphi_{000}(\vec
r_{12})>\mu_N,\eqno(6.3b)$$
$$\mu^{ex}(\Xi_c^{a0})=-\mu^{ex}(\Xi_c^{s0})=\mu^{ex}
(\Lambda_c^+)+\mu^{ex}(\Xi_c^+),\eqno(6.3c)$$
$$\mu^{ex}(\Sigma_c^{++})=0.\eqno(6.3d)$$
The exchange current corrections to the corresponding transition
magnetic moments are

$$\mu^{ex}(\Sigma_c^+\rightarrow \Lambda_c^+)={2\over
\sqrt{3}}<\varphi_{000}(\vec
r_{12})|2\tilde{V}_\pi(r_{12})-\tilde{V}_D(r_{12})|\varphi_{000}(\vec
r_{12})>,\eqno(6.4a)$$
$$\mu^{ex}(\Xi_c^{s+}\rightarrow \Xi_c^{a+})={2\over
\sqrt{3}}<\varphi_{000}(\vec
r_{12})|2\tilde{V}_K(r_{12})-\tilde{V}_{D_s}(r_{12})|\varphi_{000}
(r_{12})>,\eqno(6.4b)$$
$$\mu^{ex}(\Xi_c^{s0}\rightarrow \Xi_c^{a0})={2\over
\sqrt{3}}<\varphi_{000}(\vec r_{12})|
\tilde{V}_D(r_{12})-\tilde{V}_{D_s}(r_{12})
|\varphi_{000}(\vec
r_{12})>.\eqno(6.4c)$$
The matrix elements of the pion and kaon exchange current operators
were treated completely phenomenologically in ref. [2], and judged to
be small:\\
\noindent
 $<\varphi_{000}(\vec
r_{12})|\tilde{V}_\pi(r_{12})|\varphi_{000}(\vec r_{12})>\simeq
-0.02$, $<\varphi_{000}(\vec
r_{12})|\tilde{V}_K(r_{12})|\varphi_{000}(\vec r_{12})>\simeq 0.03$.
It is natural to expect the corresponding $D$ and $D_s$ exchange
current operators (6.2) to have even smaller matrix elements, thus 
rendering
this exchange current contribution insignificant. This implies that it
is only the transition magnetic moments (6.4) that may have
significant exchange current contributions, as it is
only these that contain
terms associated with pion and kaon exchange. In Table 6 we have
therefore only included those exchange current corrections (in column
II). These exchange current contributions are not large
enough to affect the quark model predictions of the magnetic moments
of the charm hyperons in any significant way.\\

Without any $D$ or $D_s$ exchange current contributions the magnetic
moments of the $\Lambda_c^+$, $\Xi_c^{a+}$ and $\Xi_c^{a0}$ are
predicted to be equal, the numerical value as determined by the quark
masses being $0.38\mu_N$. It is remarkable that this value almost
completely coincides with the corresponding average value $0.37\mu_N$,
which is given by the leading pion loop contribution in chiral
perturbation theory [23]. If the unknown constant in the chiral
perturbation theory calculation in ref.[23] is dropped the magnetic
moment of the $\Xi_c^{a+}$ is $0.42\mu_N$ and that of the $\Xi_c^{a0}$
is $0.37\mu_N$. Such a deviation of the charm cascade magnetic moments
from the average value can in the present approach be understood if
the $D$- and $D_s$-meson exchange current magnetic moment matrix
elements in (6.3) are negative, and the former is larger in
magntitude: e.g. $<000|\tilde{V}_D(r)|000>\sim -0.04$ and
$<000|\tilde{V}_{D_s}(r)|000>\sim -0.02$. The presence and possible
significance of charm exchange current corrections can naturally only
be decided by empirical determination of the magnetic moments of the
charm hyperons. Note that in the bound state approach to the Skyrme
model the magnetic moments of the $\Lambda_c^+$ and the
$\Xi_c^{a+},\,\,\Xi_c^{a0}$ are also predicted to be degenerate
[23].\\

In the case of the $B=-1$ hyperons the exchange current contributions
should be very small, with the exception of the transition moments,
which again also obtain a pion exchange contribution. In the case of
the $\Lambda_b$ and the $\Sigma_b$ these exchange current
contributions can be obtained from the corresponding expressions
(10.3) in ref. [2], for the strange hyperons, dropping the kaon
exchange contribution (or by replacing it by a $B$-meson exchange
contribution of the same form).\\

The impulse approximation expressions for the $B=-1$ hyperons can be
obtained from the corresponding expressions for the strange hyperons,
by substituting the $b$-quark mass in place of that of the $S$-quark.
For the $\Lambda_b^0$ and $\Sigma_b$ hyperons we obtain, 
using the quark mass values $m_u=$ 340 MeV and $m_b=5039$:

$$\mu(\Lambda_b^0)=\mu(\Xi_b^{a0})=\mu(\Xi_b^{a-})
=-{1\over 3}{m_N\over m_b}=-0.062\mu_N,\eqno(6.5a)$$
$$\mu(\Sigma_b^+)={8\over 9}{m_N\over m_u}+{1\over 9}{m_N\over
m_b}=2.47\mu_N\eqno(6.5b)$$
$$\mu(\Sigma_b^0)={2\over 9}{m_N\over m_u}+{1\over 9}{m_N\over
m_b}=0.63\mu_N\eqno(6.5c)$$
$$\mu(\Sigma^-_b)=\mu(\Omega_b^-)=-{4\over 9}{m_N\over m_u}+
{1\over 9}{m_N\over m_b}=-1.21.\eqno(6.5d)$$
The smallness of the predicted value of the magnetic moment of the
$\Lambda_b^0$ indicates that it will be difficult to measure 
accurately.\\

\vspace{1cm}
\newpage
\centerline{\bf 7. Discussion}
\vspace{0.5cm}

The results presented here show that a quite satisfactory description
of the presently known part of the spectra of the heavy flavor
hyperons can be obtained by describing the fine structure interaction
between the quarks in terms of the schematic chiral field flavor-spin
interaction (1.1), which represents the most important component of
the interaction that is mediated by the $SU(3)_F$ octet of light
pseudoscalar mesons. In order to obtain a nonvanishing splitting
between the $S=1/2$ and $S=3/2$ $C=1$ and $B=-1$ hyperons with zero
order wavefunctions with $[21]_F$ (112) and $[3]_F$ (111) flavor
symmetry, which is the analog of the octet decuplet splitting of the
light and strange baryons, a corresponding phenomenological $D$-meson
(or $D$-meson like) 
exchange interaction (1.2) also had to be included.\\

No attempt has been made here to explain the 32 MeV spin-orbit
splitting between the $\Lambda_c(2593)^+$ and $\Lambda_c(2625)^+$. It is
however 
interesting to  note
that this splitting is smaller by a factor 3.59 than the
spin-orbit splitting of 115 MeV between the corresponding strange
hyperon doublet $\Lambda(1405)-\Lambda(1520)$, because this factor
coincides almost completely with the ratio between the constituent
masses of the charm and strange hyperons $m_c/m_s=3.54$ (using the
present values for the quark masses). This strongly supports the view
that these resonances are 3-quark states, that are split by a two-body
spin-orbit interaction, which would be expected to be inversely
proportional
to the quark masses. The dynamical origin of such
a spin-orbit interaction is expected to be a combination of the
spin-orbit interaction that is associated with the effective confining
interaction and vector meson and vector meson like multimeson exchange
mechanisms [2].\\

The results presented above underpredict, as do - although to a lesser
extent - those
based on chiral perturbation
theory in ref.[14], the approximately equal
$\Sigma_c^*-\Sigma_c$ and $\Xi_c^*-\Xi_c^s$ splittings
of 75 MeV and 82
MeV respectively. The present prediction of 39 MeV for this
splitting can of course be increased to 50 - 60 MeV by relaxing
the assumption of the matrix element equality (3.2)
for the $D$ and $D_s$ exchange interaction, and the requirement
of exact reproduction of the empirical $\Sigma_c - \Lambda_c^+$
and $\Xi_c^a - \Sigma_c$
splittings. The result then would agree well with the
QCD estimate that the splitting should be approximately
$\Lambda_{QCD}^2/m_c\simeq$ 50 MeV [14]. 
If the large uncertainty limit on the preliminary 
experimental result for the $\Sigma_b^* -\Sigma_b$ splitting [20]
is taken into account, it also approximately agrees with
this quark mass scaling rule.
It would be important
that the present large empirical uncertainty limits on the
masses of the
spin $\frac{3}{2}$, $C=1$ hyperons be narrowed in order to assess
whether or not the present underprediction of the value
of the $\Sigma_c^*-\Sigma_c$ and $\Xi_c^*-\Xi_c^s$ splittings
is a problem for the present chiral quark model. The overall
quality of the predicted masses of the charm hyperons
in the ground state band is quite satisfactory, the
predictions being similar to those obtained in other
recent work using different theoretical approaches [14,23].
The quark model with the chiral field interaction (1.1)
between the light constituent quarks
is the only one, however, that at the
present time can describe the lowest negative parity resonances
of the $\Lambda_c$ in a quantitatively
satisfactory way.\\

The main conclusion of the present work is that, with the
exception of the splitting between the spin ${1\over 2}$ and 
${3\over 2}$ states in
the ground state band, the chiral field interaction (1.1) between 
the light 
constituent quarks is able to explain the empirically known part 
of the spectra of the $C=1$ hyperons. It is natural to expect the
splitting between the spin ${1\over 2}$ and 
${3\over 2}$ states in
the ground state band to reveal 
short range dynamics, which is absent or unimportant 
in the light and strange
baryons. 
This splitting may also be used to settle question of
the relative
(un)importance of the gluon exchange interaction between
heavy quarks and for pairs of light and heavy constituent quarks.
The achievement of a deeper understanding of the nature and 
the dynamical origin of this
splitting would be greatly facilitated by empirical determination of
the ground state spectrum of the $C=2$ hyperons, as the fine
structure splitting of these should be entirely due to such short
range dynamics.\\

\vspace{1cm}

{\bf Acknowledgement}
\vspace{0.5cm}

LYG thanks the Research Institute for Theoretical Physics 
at the University of Helsinki and DOR thanks the Institute for
Nuclear Theory at the University of Washington for
hospitality during the course of this work.

\newpage
\vspace{1.5cm}

{\bf References}
\vspace{0.5cm}

\begin{enumerate}
\item L.Ya. Glozman and D.O. Riska, "Systematics of the Light and
Strange Baryons and the Symmetries of QCD, Preprint HU-TFT-94-48,
(1994)
\item L.Ya. Glozman and D.O. Riska, "The Spectrum of the Nucleons and
the Strange Hyperons and Chiral Dynamics", Preprint
DOE/ER/40561-187-INT 95-16-01, Physics Reports (in press)
\item A.W. Hendry and D.B. Lichtenberg, Phys. Rev. {\bf D12}, 2756
(1975) 
\item D.B. Lichtenberg, Rep. Prog. Phys. {\bf 41}, 1707 (1978) 
\item S. Capstick and N. Isgur, Phys. Rev. {\bf D34}, 2809 (1986) 
\item A. Manohar and H. Georgi, Nucl. Phys. {\bf B234}, 189 (1984) 
\item E.J. Eichten, I. Hinchliffe and C. Quigg, Phys. Rev. {\bf D45},
2269 (1992) 
\item T.P. Cheng and L.-F. Li, Phys. Rev. Lett. {\bf 74}, 2872 (1995) 
\item J. Rosner, "Charmed Baryons with $J=3/2$", DOE/ER/40561-218,
INT95-17-05, EFI-95-48, hep-ph 9508252 (1995)
\item Particle Data Group, Phys. Rev. {\bf D50}, 1173 (1994) 
\item P. Avery et al.(CLEO collaboration), CLNS 95/1352,
CLEO 95-14, hep-ph/9508010
\item R. Werding et al. (WA89 collaboration), 27th Int. Conf.
High Energy Physics, Glasgow (1994), E. Chudakov et al., HQ94
Heavy Quarks at Fixed Targets, Virginia (1994)
\item R. Roncaglia, D. B. Lichtenberg and E. Predazzi, "Predicting
the Masses of Baryons Containing One or Two Heavy Quarks",
IUHET 293, IU/NTC 95--03, DFTT--95/7 (1995)
\item M. J. Savage, CMU-HEP95-11, DOE-ER/40682-101, hep-ph/9508268
(1995) 
\item J. Franklin, Phys. Rev. {\bf D12}, 2077 (1975) 
\item M. Rho, D.O. Riska and N.N. Scoccola, Z. Phys. {\bf A341},
343 (1992)
\item E. Jenkins and A.V. Manohar, Phys. Lett. {\bf B294}, 273
(1992) 
\item L.A. Copley, N. Isgur and G. Karl, Phys. Rev. {\bf D20},
768 (1979)
\item K. Maltman and N. Isgur, Phys. Rev. {\bf D22}, 1701 (1980) 
\item DELPHI Collaboration, "First Evidence for $\Sigma_b$ and
$\Sigma_b^*$ Baryons", DEPLHI 95-107 PHYS 542 (1995)
\item D.B. Lichtenberg, Phys. Rev. {\bf D15}, 345 (1977) 
\item L. Brekke, Ann. Phys. {\bf 240}, 400 (1995)
\item M.J. Savage, Phys. Lett. {\bf B326}, 303 (1994) 
\item Y. Oh, D.-P. Min, M. Rho and N.N. Scoccola, Nucl. Phys. {\bf
A534}, 493 (1991) 
\end{enumerate}
\newpage

\centerline{\bf Table 1}
\vspace{0.5cm}

Contributions to the masses (in MeV) 
of the $C=+1$ ground state hyperons from
flavor exchange interactions (1.1) and (1.2) ($\delta M$). The 
difference between
constituent masses of the $u$ and $s$ quarks is denoted $\Delta_s$.
The predicted mass values in column I are obtained without inclusion
of the contribution from charm exchange mechanisms. The superscripts 
$s,a$ on the $\Xi_c$ states indicate that the light and strange quarks
are in symmetric and antisymmetric states respectively. The
experimental values are from refs.[10,11,12].

\vspace{0.5cm}
\begin{center}
\begin{tabular}{|l|l|l|l|l|} \hline
$[f]_{FS}[f]_{F} [f]_S$ & State & Predicted & Predicted & $\delta M$ \\
 & (mass) & mass I & mass II & \\ \hline
$[3]_{FS}[21]_{F}[21]_S$ & $\Lambda_c^+$ & 2285 & 2285 &
$-9P_{00}^\pi+P_{00}^{uu}-6P_{00}^{D}$ \\
 & (2285) & (input) & (input) & $-{1\over 2}\delta$\\ \hline
$[3]_{FS}[21]_{F}[21]_S$ & $\Sigma_c^+$ & 2481 & 2455 & $-P_{00}^\pi-{1\over
3}P_{00}^{uu}-10P_{00}^{D}$ \\
 & (2455?) & & (input) & $-{1\over 2}\delta$\\
$[3]_{FS}[3]_{F} [3]_S$ & $\Sigma_c^*$ & 2481 & 2494 & $-P_{00}^\pi-{1\over
3}P_{00}^{uu}-4P_{00}^{D}$ \\
 & (2530?) & & & $-{1\over 2} \delta$\\ \hline
$[3]_{FS}[21]_{F}[21]_S$ & $\Xi_c^{(a)}$ & 2485 & 2485 &
$-6P_{00}^K-2P_{00}^{us}$ \\ 
 & (2465-2470) & & &$-3P_{00}^{D}-3P_{00}^{D_s}$ \\
 & & & & $+\Delta_s-{1\over 2}\delta$ \\
$[3]_{FS}[21]_{F}[21]_S$ & $\Xi_c^{(s)}$ & 2618 & 2593 &
$-2P_{00}^{K}+{2\over 3} P_{00}^{us}$ \\
 & (2560?)& & & $-5P_{00}^{D}-5P_{00}^{D_s}$ \\
 & & & & $+\Delta_s-{1\over 2}\delta$ \\
$[3]_{FS}[3]_{F}[3]_S$ & $\Xi_c^*$ & 2618 & 2632 & $-2P_{00}^{K}+{2\over
3}P_{00}^{us}$ \\
 & (2642?)& & & $-2P_{00}^{D}-2P_{00}^{D_s}$ \\
 & & & & $+\Delta_s-{1\over 2}\delta$ \\ \hline
$[3]_{FS}[21]_{F}[21]_s$ & $\Omega_c^0$ & 2748 & 2726 & $-{4\over
3}P_{00}^{ss}-10P_{00}^{D_s}$ \\
 & (2710?) & & & $+2\Delta_s-{1\over 2}\delta$ \\
$[3]_{FS}[3]_{F}[3]_S$ & $\Omega_c^{0^*}$ & 2748 & 2765 & $-{4\over
3}P_{00}^{ss}-4P_{00}^{D_s}$ \\
 & & & & $+2\Delta_s-{1\over 2}\delta$ \\ \hline
\end{tabular}
\end{center}

\newpage

\centerline{\bf Table 2}
\vspace{0.5cm}

The negative parity $L=1$ $\Lambda_c^+$ and $\Sigma_c$ resonances as
predicted in the harmonic oscillator model with the flavor exchange
fine structure interactions (1.1) and (1.2). The fine structure
corrections include the $u-s$ quark mass difference $\Delta_s$ and the
corrections $\delta$ due to the mass difference between the light
and heavy quarks (4.1). The predicted energies (in MeV)
are given in brackets.
The predicted value for the lowest $\Lambda_c^+$ doublet in square
brackets is obtained with $P_{11}^{D} = 0$.  

\vspace{0.5cm}
\begin{center}
\begin{tabular}{|l|l|l|l|} \hline
$[f]_{FS}[f]_F[f]_S$ & Multiplet & average & $\delta M$ \\ 
 & & energy &  \\ \hline
$[21]_{FS}[111]_F[21]_S$ & ${1\over 2}^-,\Lambda_c (2593)^+?$ & 2609 &
$-{9\over 2}P_{00}^\pi+{1\over 2}P_{00}^{uu}-6P_{00}^{D}$\\
 & ${3\over 2}^-,\Lambda_c(2625)^+?$ & (2609) & $+{3\over
2}P_{11}^\pi-{1\over 6}P_{11}^{uu}+2P_{11}^{D}$\\
 & &[2599] & $-\frac{2}{3}\delta$ \\ \hline
$[21]_{FS}[21]_F[21]_S$ & ${1\over 2}^-,\Lambda_c^+;$ & ? & $-{9\over
2}P_{00}^\pi+{1\over 2}P_{00}^{uu}-3P_{00}^{D}$\\
 &${3\over 2}^-,\Lambda_c^+$
 &(2643) & $+{3\over 2}P_{11}^\pi-{1\over 6}P_{11}^{uu}+5P_{11}^{D}$ \\
 &  &  & $-\frac{2}{3}\delta$ 
\\ \hline
$[21]_{FS}[21]_F[3]_S$ & ${1\over 2}^-,\Lambda_c^+;$
 & ? & $-3P_{00}^{D}$ \\
 & ${3\over
2}^-,\Lambda_c^+;$ &(2655)
 & $+3P_{11}^\pi-{1\over 3}P_{11}^{uu}+P_{11}^{D}$\\
 & ${5\over 2}^-,\Lambda_c^+$ & & $-\frac{7}{12}\delta
$ \\ \hline
$[21]_{FS}[21]_F[21]_S$ & ${1\over 2}^-,\Sigma_c$; & ? & $-{1\over
2}P_{00}^\pi-{1\over 6}P_{00}^{uu}-5P_{00}^{D}$ \\
 & ${3\over 2}^-,\Sigma_c$ & (2747) & $+{3\over
2}P_{11}^\pi+{1\over 2}P_{11}^{uu}+3P_{11}^{D}$\\
 & & & $-{2\over 3}\delta$ \\ \hline
$[21]_{FS}[3]_F[21]_S$ & ${1\over 2}^-,\Sigma_c$; & ? & $-{1\over
2}P_{00}^\pi-{1\over 6}P_{00}^{uu}-2P_{00}^{D}$\\
 & ${3\over 2}^-,\Sigma_c$
 & (2693)  & $+{3\over 2}P_{11}^\pi
+{1\over 2}P_{11}^{uu}+6P_{11}^{DD^*}$ \\
 & && $-{2\over 3}\delta$ \\ \hline
$[21]_{FS}[21]_F[3]_S$ & ${1\over 2}^-,\Sigma_c$; & ? &
$-P_{00}^\pi-{1\over 3}P_{00}^{uu}-P_{00}^{D}$\\
 & ${3\over 2}^-,\Sigma_c$; & (2654)  & $+3P_{11}^{D}$\\
 & ${5\over 2}^-,\Sigma_c$ && $-{3\over 4}\delta$\\ \hline
\end{tabular}
\end{center}

\newpage

\centerline{\bf Table 3}
\vspace{0.5cm}

The negative parity $L=1$ $\Xi_c$ and $\Omega_c^0$
resonances predicted with the flavor
exchange fine structure interactions (1.1) and (1.2). The quark mass
difference corrections $\Delta_s$ and $\delta$ are indicated
explicitly. The energies are given in units of MeV.
\vspace{0.5cm} 
\begin{center}
\begin{tabular}{|l|l|l|l|} \hline
$[f]_{FS}[f]_F[f]_S$ & Multiplet & average & $\delta M$\\
 & & energy & \\ \hline
$[21]_{FS}[111]_F[21]_S$ & ${1\over 2}^-,\Xi_c^a$; & 2752 &
$-P_{00}^{us}-3P_{00}^K-3 P_{00}^{D}-3P_{00}^{D_s}$ \\
 & ${3\over 2}^-, \Xi_c^a$ &  & $+{1\over
3}P_{11}^{us}+P_{11}^K+P_{11}^{D} +P_{11}^{D_s}$ \\
 & & & $+\Delta_s-{2\over 3}\delta$ \\ \hline
$[21]_{FS}[21]_F[21]_S$ & ${1\over 2}^-, \Xi_c^a$; & 2787 & $
-P_{00}^{us}-3P_{00}^K-{3\over 2}P_{00}^{D}-{3\over
2}P_{00}^{D_s}$ \\
 & ${3\over 2}^-,\Xi_c^a$ & & $+{1\over 3}P_{00}^{us}+P_{11}^K
+{5\over 2}P_{11}^{D}+{5\over 2}D_{11}^{D_s}$ \\
 & & & $+\Delta_s-{2\over 3}\delta$\\ \hline
$[21]_{FS}[21]_F[3]_S$ & ${1\over 2}^-,\Xi_c^a; $
 & 2898 & $-{3\over 2}P_{00}^{D}-{3\over
2}P_{00}^{D_s}$\\
 & ${3\over
2}^-,\Xi_c^a;$ & & $+{2\over 3}P_{11}^{us}+2P_{11}^K+
{1\over
2}P_{11}^{D}+{1\over 2}P_{11}^{D_s}$\\
 &${5\over 2}^-,\Xi_c^a$ &  & $+\Delta_s-{7\over 12}\delta$\\ \hline
$[21]_{FS}[21]_F[21]_S$ & ${1\over 2}^-,\Xi_c^s;$ & 2851 & ${1\over
3}P_{00}^{us}-P_{00}^K-{5\over 2}P_{00}^{D}-{5\over
2}P_{00}^{D_s}$\\
 & ${3\over 2}^-,\Xi_c^s$ &  & $3P_{11}^K-P_{11}^{us}
+{3\over2}P_{11}^{D}+{3\over 2}P_{11}^{D_s}$ \\
 & & & $+\Delta_s-{2\over 3}\delta$ \\ \hline
$[21]_{FS}[3]_F[21]_S$ & ${1\over 2}^-,\Xi_c^s;$ & 2886 & ${1\over
3}P_{00}^{us}-P_{00}^K-P_{00}^{D}-P_{00}^{D_s}$\\
 & ${3\over 2}^-,\Xi_c^s$ &  &
$-P_{11}^{us}+3P_{11}^K+3P_{11}^{D}+3P_{11}^{D_s}$ \\
 & & & $+\Delta_s-{2\over 3}\delta$ \\ \hline
$[21]_{FS}[21]_F[3]_S$ & ${1\over 2}^-,\Xi_c^s;$ & 2792 & ${2\over
3}P_{00}^{us}-2P_{00}^{K}-{1\over 2}P_{00}^{D}-{1\over
2}P_{00}^{D_s}$ \\
 & ${3\over 2}^-,\Xi_c^s;$ & & $+{3\over
2}P_{11}^{D}+{3\over 2}P_{11}^{D_s}$\\
 & ${5\over 2}^-,\Xi_c^s$ & & $+\Delta_s-{3\over 4}\delta$\\ \hline
$[21]_{FS}[21]_F[21]_S$ & ${1\over 2}^-,\Omega_c^0;
$ & 2965 & $-{2\over
3}P_{00}^{ss}-5P_{00}^{D_s}$\\
 & ${3\over 2}^-,\Omega_c^0$ & & $+2P_{11}^{ss}
+3P_{11}^{D_s}$ \\
 & & & $+2\Delta_s-{2\over 3}\delta$ \\ \hline
$[21]_{FS}[3]_F[21]_S$ & ${1\over 2}^-,\Omega_c^0;$ & 2999 & $-{2\over
3}P_{00}^{ss}-2P_{00}^{D_s}$\\
 & ${3\over 2}^-,\Omega_c^0$ &  &
$+2P_{11}^{ss}+6P_{11}^{D_s}$ \\
 & & & $+2\Delta_s-{2\over 3}\delta$ \\ \hline
$[21]_{FS}[21]_F[3]_S$ & ${1\over 2}^-,\Omega_c^0;$ & 2936 & $-{4\over
3}P_{00}^{ss}-P_{00}^{D_s}$ \\
 & ${3\over 2}^-,\Omega_c^0;$ &  & $+3P_{11}^{D_s}$\\
 & ${5\over 2}^-,\Omega_c^0$ & & $+2\Delta_s-{3\over 4}\delta$\\ \hline
\end{tabular}
\end{center}
\newpage
\centerline{\bf Table 4}
\vspace{0.5cm}

Contributions to the masses (in MeV) 
of the $B=-1$ ground state hyperons from
flavor exchange interactions (1.1) and (1.2) ($\delta M$). The 
difference between
constituent masses of the $u$ and $s$ quarks is denoted $\Delta_s$.
The predicted mass values  are given in brackets. 
The superscripts 
$s,a$ on the $\Xi_b$ states indicate that the light and strange quarks
are in symmetric and antisymmetric states respectively. The empirical
values are from refs. [10,20]. 

\vspace{0.5cm}
\begin{center}
\begin{tabular}{|l|l|l|l|l|} \hline
$[f]_{FX} [f]_S$ & State & Mass &  $\delta M$ \\
 & (mass)  & & \\ \hline
$[3]_{FS}[21]_{F}[21]_S$ & $\Lambda_b^0$ & 5641  &
$-9P_{00}^\pi+P_{00}^{uu}-6P_{00}^B$ \\
 &  & (input) & $-{1\over 2}\delta$\\ \hline
$[3]_{FS}[21]_{F}[21]_S$ & $\Sigma_b$ & 5814 & $-P_{00}^\pi-{1\over
3}P_{00}^{uu}-10 P_{00}^B$ \\
&&(input)&$-{1\over 2}\delta$\\
$[3]_{FS}[3]_{F} [3]_S$ & $\Sigma_b^*$  & 5870 & 
$-P_{00}^\pi-{1\over
3}P_{00}^{uu}-4 P_{00}^B$\\
 & &(5850)  & $-{1\over 2}\delta$\\
 \hline
$[3]_{FS}[21]_{F}[21]_S$ & $\Xi_b^{(a)}$ & ? &
$-6P_{00}^K-2P_{00}^{us}$ \\ 
 & & (5842) & $-3P_{00}^B-3P_{00}^{B_s}$\\ 
&&& $+\Delta_s-{1\over 2}\delta$\\
$[3]_{FS}[21]_{F}[21]_S$ & $\Xi_b^{(s)}$ &?  &
$-2P_{00}^{K}+{2\over 3} P_{00}^{us}$ \\
&&(5952)&$-5P_{00}^B-5P_{00}^{B_s}$\\
&&& $+\Delta_s-{1\over 2}\delta$\\
$[3]_{FS}[3]_{F}[3]_S$ & $\Xi_b^*$  &? & $-2P_{00}^{K}
+{2\over 3} P_{00}^{us}$ \\
&&(5988)&$-2P_{00}^B-2P_{00}^{B_s}$\\
&&&  $+\Delta_s-{1\over 2}\delta$\\
 \hline
$[3]_{FS}[21]_{F}[21]_S$ & $\Omega_b^0$  &?  & $-{4\over
3}P_{00}^{ss}-10P_{00}^{B_s}$ \\
&&(6080)& $+2\Delta_s-
\frac{1}{2}\delta$\\
$[3]_{FS}[3]_{F}[3]_S$ &$\Omega_b^{0^*}$ &?&$-{4\over
3}P_{00}^{ss}-4P_{00}^{B_s}$\\
&&(6116)&$+2\Delta_s-\frac{1}{2}\delta$ \\
\hline
\end{tabular}
\end{center}

\newpage

\centerline{\bf Table 5}
\vspace{0.5cm}

The negative parity $L=1$ $B=-1$ hyperon resonance energies (in MeV) as
predicted in the harmonic oscillator model with the flavor exchange
fine structure interactions (1.1). The fine structure
corrections include the $u-s$ quark mass difference $\Delta_s$ and the
corrections $\delta$ due to the mass difference between the light
and heavy quarks (4.1). The neglect of the corrections of the
$B-$ and $B_s -$ exchange interactions implies a 20--30 MeV
uncertainty in the predicted energies.

\vspace{0.5cm}
\begin{center}
\begin{tabular}{|l|l|l|l|} \hline
$[f]_{FS}[f]_F[f]_S$ & Multiplet & Predicted & $\delta M$ \\ 
 & & energy &  \\ \hline
$[21]_{FS}[111]_F[21]_S$ & ${1\over 2}^-,\Lambda_b^0; $ & 5940 &
$-{9\over 2}P_{00}^\pi+{1\over 2}P_{00}^{uu}+{3\over
2}P_{11}^\pi$\\
$[21]_{FS}[21]_F[21]_S$ & 
${3\over 2}^-,\Lambda_b^0$ 
&  & $-{1\over 6}P_{11}^{uu}-\frac{2}{3}\delta$\\ \hline
$[21]_{FS}[21]_F[3]_S$ & ${1\over 2}^-,\Lambda_b^0;
{3\over 2}^-,\Lambda_b^0; $
 & 6150 &$+3P_{11}^\pi-{1\over 3}P_{11}^{uu}$  \\
 & ${5\over
2}^-,\Lambda_b^0;$ &
 & $-\frac{7}{12}\delta
$ \\
 \hline
$[21]_{FS}[21]_F[21]_S$ & ${1\over 2}^-,\Sigma_b$; & 6070 & $-{1\over
2}P_{00}^\pi-{1\over 6}P_{00}^{uu}+{3\over
2}P_{11}^\pi$ \\
$[21]_{FS}[3]_F[21]_S$ & ${3\over 2}^-,\Sigma_b$ &  &
$+{1\over 2}P_{11}^{uu}-{2\over 3}\delta$ \\
\hline
$[21]_{FS}[21]_F[3]_S$ & ${1\over 2}^-,\Sigma_b; 
{3\over 2}^-,\Sigma_b;$ & 5950 &
$-P_{00}^\pi-{1\over 3}P_{00}^{uu}$\\
 & ${5\over 2}^-,\Sigma_b$; &   &$-{3\over 4}\delta$ \\
 \hline
$[21]_{FS}[111]_F[21]_S$ & ${1\over 2}^-,\Xi_b^a$; & 6110 &
$-P_{00}^{us}-3P_{00}^K+P_{11}^K$ \\
$[21]_{FS}[21]_F[21]_S$ 
& ${3\over 2}^-, \Xi_b^a$ &  &$+{1\over
3}P_{11}^{us}+\Delta_s-{2\over 3}\delta$  \\
 \hline
$[21]_{FS}[21]_F[3]_S$ & ${1\over 2}^-,\Xi_b^a; 
{3\over 2}^-,\Xi_c^a;$
 & 6240 & $+{2\over 3}P_{11}^{us}+2P_{11}^K
$\\
 & ${5\over
2}^-,\Xi_b^a;$ & & $+\Delta_s-{7\over 12}\delta$\\
 \hline
$[21]_{FS}[21]_F[21]_S$ & ${1\over 2}^-,\Xi_b^s;$ & 6200 & ${1\over
3}P_{00}^{us}-P_{00}^K +3P_{11}^K$\\
$[21]_{FS}[3]_F[21]_S$  
& ${3\over 2}^-,\Xi_b^s$ &  & $-P_{11}^{us}+\Delta_s
-{2\over 3}\delta$
 \\
 \hline
$[21]_{FS}[21]_F[3]_S$ & ${1\over 2}^-,\Xi_b^s;
{3\over 2}^-,\Xi_c^s;$ & 6120 & ${2\over
3}P_{00}^{us}-2P_{00}^{K}$ \\
 & ${5\over 2}^-,\Xi_b^s$ &  &$+\Delta_s-{3\over 4}\delta$ \\
 \hline
$[21]_{FS}[21]_F[21]_S$ & ${1\over 2}^-,\Omega_b^-;$ & 6310 & $-{2\over
3}P_{00}^{ss}+2P_{11}^{ss}$\\
$[21]_{FS}[3]_F[21]_S$
 & ${3\over 2}^-,\Omega_b^-$ &  &$+2\Delta_s-{2\over 3}\delta$  \\
 \hline
$[21]_{FS}[21]_F[3]_S$ & ${1\over 2}^-,\Omega_b^-;
{3\over 2}^-,\Omega_b^-;$ & 6240 & $-{4\over
3}P_{00}^{ss}$ \\
 & ${5\over 2}^-,\Omega_b^-$ &  &$+2\Delta_s-{3\over 4}\delta$\\
 \hline
\end{tabular}
\end{center}
\newpage

\centerline{\bf Table 6}
\vspace{0.5cm}

Magnetic moments of the $S=1/2$ $C=+1$ charm hyperons in the quark
model (in units of the nuclear magneton). 
Column IA contains the impulse approximation expressions and
column I the corresponding numerical values. Column II contains
the exchange current corrections and column III the combined net
prediction.

\vspace{0.5cm}
\begin{center}
\begin{tabular}{|l|l|r|r|r|} \hline
 & IA & I & II & III \\ \hline
$\Lambda_c^+$ & ${2\over 3} {m_N\over m_c}$ & 0.38 & & 0.38 \\
$\Sigma_c^{++}$ & ${8\over 9}{m_N\over m_u}-{2\over 9}{m_N\over m_c}$
& 2.33 & & 2.33\\
$\Sigma_c^+$ & ${2\over 9}{m_N\over m_u}-{2\over 9}{m_N\over m_c}$ &
0.49 & & 0.49\\
$\Sigma_c^0$ & $-{4\over 9} {m_N\over m_u}-{2\over 9}{m_N\over m_c}$ &
--1.35 & & --1.35\\
$\Sigma_c^+\rightarrow \Lambda_c^+$ & $-{1\over \sqrt{3}}{m_N\over
m_u}$ & --1.59 & --0.04 & --1.63\\
$\Xi_c^{a+}$ & ${2\over 3}{m_N\over m_c}$ & 0.38 & & 0.38\\
$\Xi_c^{a0}$ & ${2\over 3}{m_N\over m_c}$ & 0.38 & & 0.38\\
$\Xi_c^{s+}$ & ${4\over 9}{m_N\over m_u}-{2\over 9}{m_N\over
m_s}-{2\over 9}{m_N\over m_c}$ & 0.65 & & 0.65\\
$\Xi_c^{s0}$ & $-{2\over 9}{m_N\over m_u}-{2\over 9}{m_N\over
m_s}-{2\over 9}{m_N\over m_c}$ & --1.18 & & --1.18\\
$\Xi_c^{s+}\rightarrow \Xi_c^{a+}$ & $-{1\over \sqrt{3}}(2{m_N\over
m_u}+{m_N\over m_s})$ & --1.45 & 0.07 & --1.38\\
$\Xi_c^{s0}\rightarrow \Xi_c^{a0}$ & ${1\over 3\sqrt{3}}({m_N\over
m_u}-{m_N\over m_s})$ & 0.14 &  & 0.14\\
$\Omega_c^0$ & $-{4\over 9}{m_N\over m_s}-{2\over 9}{m_N\over m_c}$ &
--1.02 & & --1.02\\ \hline
\end{tabular}
\end{center}

\end{document}